\documentclass[review]{elsarticle}

\usepackage{graphicx}
\usepackage{wrapfig}

\begin{document}

\begin{frontmatter}

\title{Phase transformation in boron under shock compression}
\tnotetext[mytitlenote]{LLNL-JRNL-803936-DRAFT}

\author[myaddress1,myaddress2]{Shuai Zhang\fnref{myfootnote1}}
\fntext[myfootnote1]{Corresponding author}
\ead{szha@lle.rochester.edu}

\author[myaddress2]{Heather D. Whitley}

\author[myaddress2]{Tadashi Ogitsu}

\address[myaddress1]{Laboratory for Laser Energetics, University of Rochester, Rochester, New York 14623, USA}
\address[myaddress2]{Lawrence Livermore National Laboratory, Livermore, California 94550, USA}

\begin{abstract}
\begin{wrapfigure}[11]{l}{6.5cm}
\includegraphics[width=6.5cm]{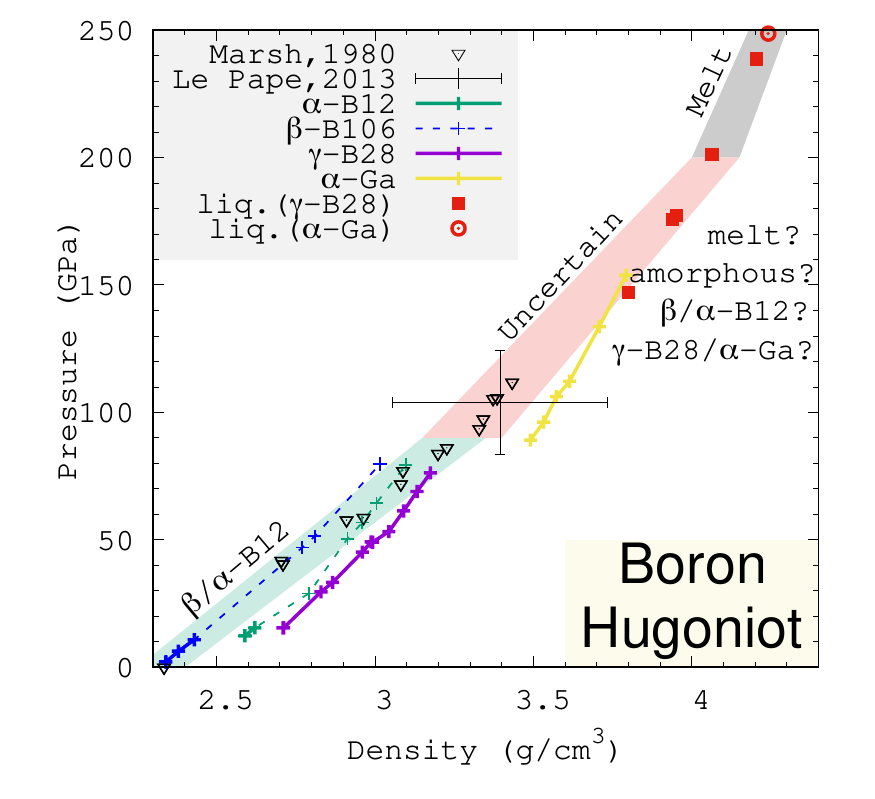}
\end{wrapfigure}
Using first-principles molecular dynamics, we calculated the equation of state and shock Hugoniot of various boron phases. 
We find a large mismatch between Hugoniots based on existing knowledge of the equilibrium phase diagram and those measured by shock experiments, which could be reconciled if the $\alpha$-B$_{12}$/$\beta \rightarrow\gamma$-B$_{28}$ transition is significantly over-pressurized in boron under shock compression. Our results also indicate that
there exists an anomaly and negative Clapeyron slope along the melting curve of boron at 100 GPa and 1500--3000 Kelvin.
These results enable in-depth understanding of matter under shock compression, in particular the significance of compression-rate dependence of phase transitions and kinetic effects in experimental measurements. 
\end{abstract}


\end{frontmatter}


\section{Introduction}

Boron is a prototype for low-$Z$ and superhard materials and a candidate for making ablators for high energy density and inertial confinement fusion experiments~\cite{Zhang2018b}. 
Boron has complex atomistic structures that are characterized by various icosahedral arrangements at ambient conditions and multiple phases under moderate (up to $\sim$3.5-times) compression.
Clarifying the structure, stability relation, and melting of the various boron polymorphs at high pressure has long been a subject of interest in the solid state physics and chemistry community~\cite{Masago2006,Shang2007,vanSetten2007,White2015,Ogitsu2009jacs,Ogitsu2013,Ma2003,Zarechnaya2009prl,Parakhonskiy2011,Kurakevych2012,Qin2012,Solozhenko2013,SanzPRL2002amphous,ZhangPRM2018gammab28}.

Existing phase diagrams of boron are represented by five different crystalline phases: $\alpha,\beta,\gamma$, t, and $\alpha$-Ga, according to density functional theory (DFT) calculations and diamond anvil cell (DAC) experiments at pressures up to $\sim$200 GPa~\cite{Oganov2009}.
Among them, the nonicosahedral metallic $\alpha$-Ga phase has not been experimentally verified until recently~\cite{Chuvashova2017}, and
questions remain on the exact boundaries between different phases and the atomistic structure of the t phase~\cite{Ogitsu2013,Parakhonskiy2011,Shirai2017}.
Dynamic compression experiments, which  simultaneously generate high pressures and high temperatures, produced pressure-density equation of state (EOS) data along the shock Hugoniot of solid boron up to 112 GPa~\cite{MarshLASL1980}. More recently, additional data have been obtained from experiments at large laser facilities, including liquid structure-factor measurements using x-ray radiography to 100--400 GPa~\cite{lepape2013} and warm dense matter EOS measurements of the Hugoniot at pressures as high as 5608 GPa~\cite{Zhang2018b}. Explosive shock experiments have also been performed to measure the electrical conductivity~\cite{Molodets2017} and structural transformation using x-ray diffraction~\cite{Molodets2018} and found amorphization of $\beta$ boron when shock-compressed to 115 GPa.

These recent laser-shock results~\cite{Zhang2018b,lepape2013} have been compared with highly accurate first-principles electronic structure methods, in particular density functional theory molecular dynamics (DFT-MD).  The DFT-MD calculations have shown remarkably good agreement with experimental results for the ionic structure and Hugoniot of liquid boron, and also provided the opportunity to examine other details of the material that are not immediately accessible in shock experiments, such as the Hugoniot temperature and diffusivity.

In this paper, we focus on the condensed matter regime of the phase diagram with pressures up to 1000 GPa and temperatures up to a few electron volt (eV, 1~eV$\approx$1.2$\times10^4$~K). These conditions are accessible with routine diamond anvil cell and shock experiments, which can be used to test our predictions.



\section{Methodology}
Our DFT-MD simulations of boron follow similar procedures to those in~\cite{Zhang2018b}. We use Vienna \textit{Ab initio} Simulation Package
({\footnotesize VASP})~\cite{kresse96b} for all the calculations. We choose a hard (core radius=1.1 Bohr) projected augmented wave (PAW)
pseudopotential~\cite{Blochl1994} which is suitable for high-pressure studies.
The 2s$^2$2p$^1$ electrons are treated as valence.
Perdew-Burke-Ernzerhof (PBE)~\cite{Perdew96} exchange-correlation functional is used in this work,
consistent with previous DFT calculations on solid boron~\cite{Ogitsu2010,Prasad2005,Siberchicot2009}.
We use a large cutoff energy of 2000 eV for the plane-wave basis and the
$\Gamma$ point for Brillouin zone sampling. 
MD trajectories, typically consisting of 5000--10000 snapshots each with timesteps of 0.05--0.55 fs and regulated by the Nos\'{e} thermostat~\cite{Nose1984}, are generated
to form canonical ($NVT$) ensembles.
Internal energies and pressures are obtained by averaging over the trajectory following equilibration of the ionic structure, typically after the first 20\% of the simulation time. 
The atomistic diffusion during the DFT-MD simulations have been carefully cross-checked to ensure no structural instability occurs at the conditions that the EOS data are reported.

For solid boron, we refer to previous publications for the equilibrium phase diagram~\cite{Solozhenko2013,Oganov2009} and perform DFT-MD calculations for four major phases ($\beta$, $\alpha$-B$_{12}$, $\gamma$-B$_{28}$, and $\alpha$-Ga) at pressure-temperature conditions beyond their respective stability regimes. This allows construction of Hugoniots in broader ranges, which enables more detailed comparison with experiments than being limited by the phase boundaries and their uncertainties in the equilibrium phase diagram. In the DFT-MD simulations, we choose a 2$\times$2$\times$2 supercell with 96 atoms for $\alpha$-B$_{12}$, 3$\times$2$\times$3 supercell with 144 atoms for $\alpha$-Ga, 2$\times$2$\times$1 supercell with 112 atoms for $\gamma$-B$_{28}$, and two different structures for $\beta$ boron~\cite{AnPRL2016B}: one is symmetric with 105 atoms in the unit cell, whereas the other has 106 atoms in the unit cell.

The simulations of liquid boron typically consist of 120 atoms in a cubic box with periodic boundary conditions. With this setting, our previous tests with different cell sizes~\cite{Zhang2018b} showed that the finite size errors on the EOS are negligible at 5$\times10^4$~K. We perform additional calculations at 1500--2$\times10^4$ K using the same cell shapes and numbers of atoms as the corresponding solid phases, in order to examine the possibility of structure dependence of the EOS of boron liquids.

\begin{figure}
\centering\includegraphics[width=0.49\textwidth]{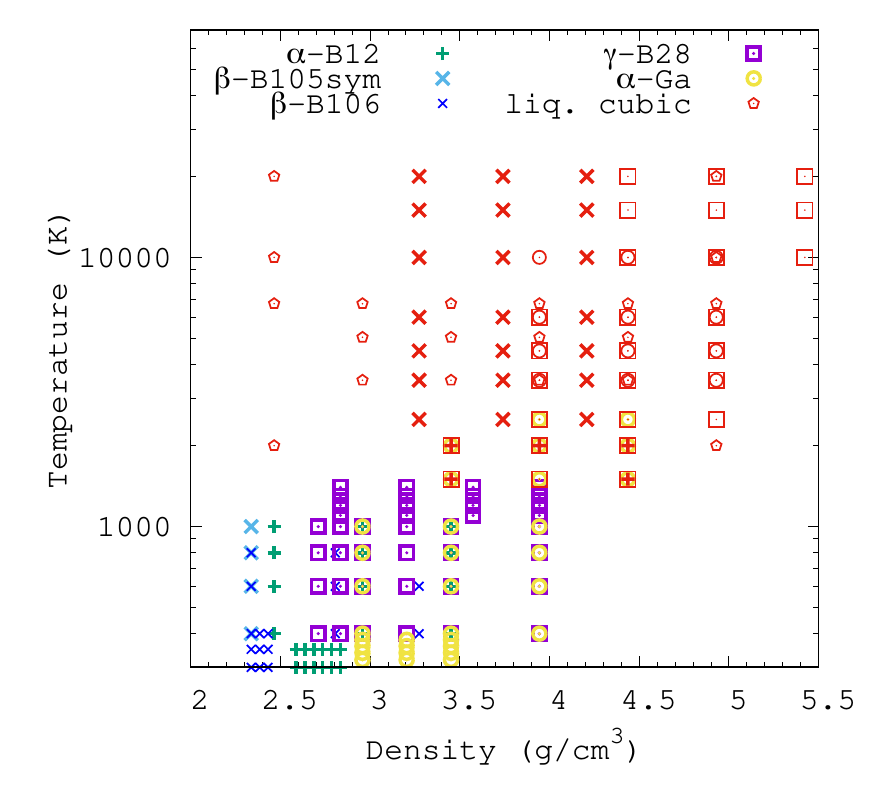}
\centering\includegraphics[width=0.49\textwidth]{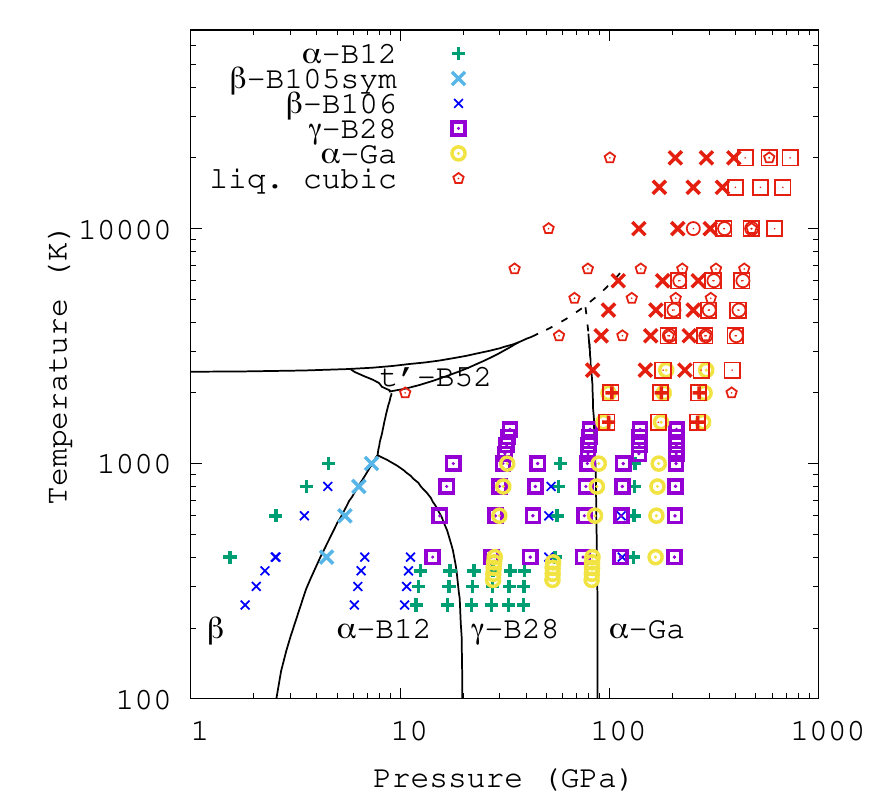}
\caption{\label{fig:eosgrid} Boron EOS data points from this work shown in temperature-density (left) and temperature-pressure (right) diagrams. Colored symbols denote different solid phases or liquids (in red color) that are simulated with the same cell shape as the corresponding solid. Phase boundaries based on previous literature~\cite{Solozhenko2013,Oganov2009} are shown with solid black curves. The dashed black curves are extrapolation of the melting curve and the $\gamma$-B$_{28}$$\leftrightarrow$$\alpha$-Ga boundary using three-parameter Kechin's equation~\cite{Kechin2001}. The melting curve is anchored at 2440 K and 0 GPa and the values of the parameters are $a$=2.3~GPa, $b$=-0.01, and $c$=-0.009~GPa$^{-1}$. The $\gamma$-B$_{28}$$\leftrightarrow$$\alpha$-Ga phase boundary is anchored at 7.3 K and 88.4 GPa and the values of the parameters are $a$=-0.023~GPa, $b$=1.04, and $c$=-0.0004~GPa$^{-1}$.}
\end{figure}

\section{Results and Discussion}

Our DFT-MD simulations and EOS data for boron reveal several  facts about the structural relations and the phase diagram of boron, as shown in Fig.~\ref{fig:eosgrid}, 
which summarizes the temperature-density conditions where the explored phases can be stabilized in this study and the corresponding temperature-pressure data. 
Firstly, along two isochores of 2.96 and 3.45~g/cm$^3$, the pressure relation of the three $\alpha$-B$_{12}$, $\gamma$-B$_{28}$, and $\alpha$-Ga phases are $P^{\alpha\textrm{-Ga}} < P^{\gamma\textrm{-B}_{28}} < P^{\alpha\textrm{-B}_{12}}$. This indicates that the transitions $\alpha$-B$_{12}$ $\rightarrow\gamma$-B$_{28}$ $\rightarrow\alpha$-Ga are driven by the $PV$ term in the enthalpy differences because the density increases discontinuously upon transition to the high-pressure phases. 
Secondly, along the isochore of 2.34~g/cm$^3$, the pressures of the symmetric 105-atom structure is higher than the 106-atom $\beta$ phase. This, together with our findings that the symmetric 105-atom structure is not stable at higher densities, indicates that the structural complexity of the $\beta$ phase can have more significant effects in determining the properties of boron at ambient--low pressures than at high pressures.
Thirdly, our simulations show that the melting temperature of high-pressure boron can be as low as 1500--2500~K (for the $\alpha$-B$_{12}$, $\beta$, and $\gamma$ phases) or 2500--3500~K (for the $\alpha$-Ga phase) at above 90 GPa. This is in contrast to expectations based on existing knowledge about the boron phase diagram (dashed curves in the right panel of Fig.~\ref{fig:eosgrid}) and indicates that an anomaly and negative Clapeyron slope could exist along the melting curve at 100 GPa and 1500--3000 K.

The above results and discussion are about the equilibrium phases of boron under stable temperatures and pressures. In the following, we focus on phase transitions under shock compression.
The EOS data from our DFT-MD simulations are utilized to predict the shock Hugoniot profiles, following the Rankine-Hugoniot equation~\cite{Meyers1994book}
$(E-E_0) + (P+P_0)(V-V_0)/2 = 0$
which connects the internal energy, pressure, and volume of the shocked state ($E,P,V$) to the initial state ($E_0,P_0,V_0$).
The initial energy $E_0$ and pressure $P_0$ are taken to be those of $\beta$ boron with a density of 2.338~g/cm$^3$ according to DFT-MD simulations of a 1280-atom structure at 300~K~\cite{Ogitsu2009jacs,Ogitsu2010,initrhonote}.

\begin{figure}
\centering\includegraphics[width=0.98\textwidth]{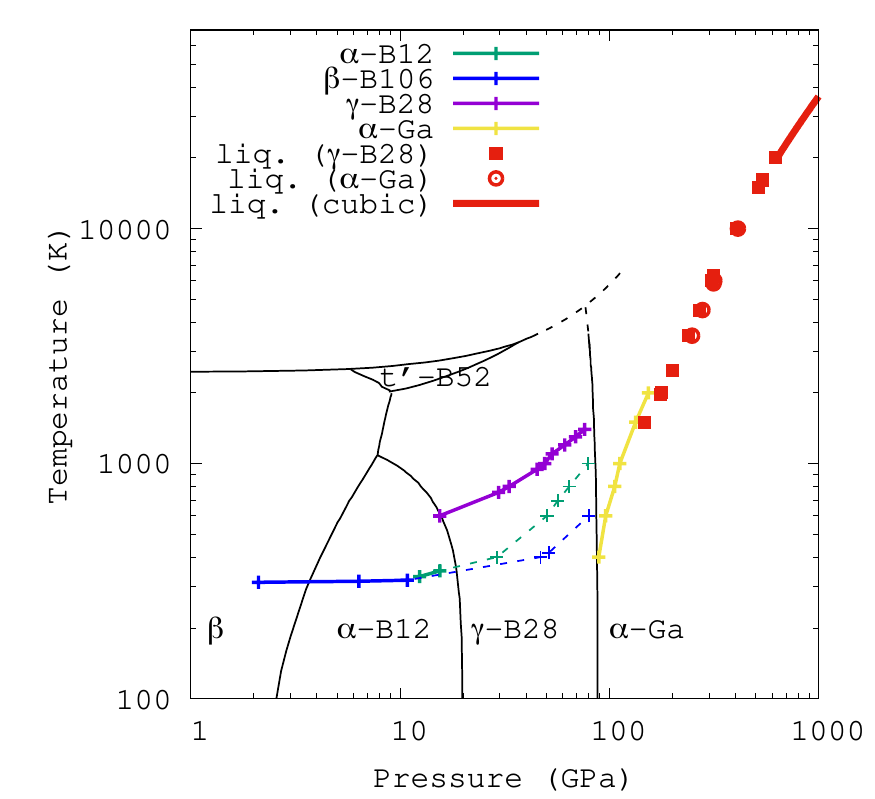}
\caption{\label{fig:hugTP} Hugoniots of various boron phases plotted in the equilibrium phase diagram based on previous literature~\cite{Solozhenko2013,Oganov2009}. The dashed colored curves are expected Hugoniot profiles of $\beta$ and $\alpha$-B$_{12}$ boron if the sample is shocked to the corresponding pressures but does not transform to other phases. The lines are guides to the eyes.}
\end{figure}

\begin{figure}
\centering\includegraphics[width=0.98\textwidth]{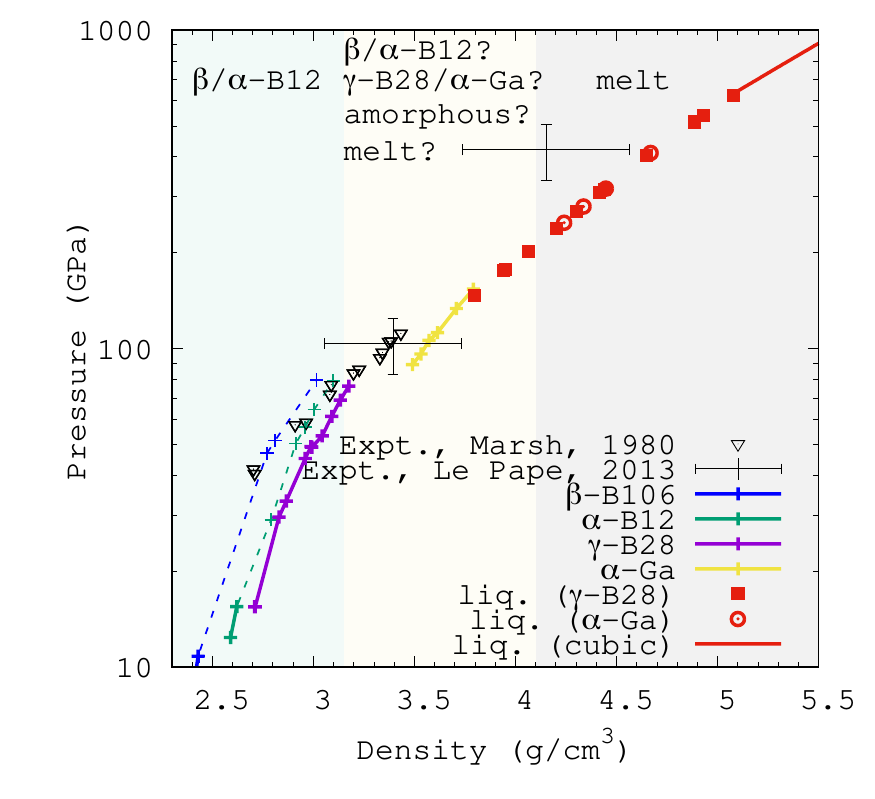}
\caption{\label{fig:hugPr} Pressure-density Hugoniots of various boron phases obtained from the present work in comparison with experimental data~\cite{MarshLASL1980,lepape2013}. Solid lines denote the principal Hugoniot of $\beta$ boron if assuming phase boundaries follow the equilibrium diagram shown in Fig.~\ref{fig:hugTP}.~\cite{hugprnote1,hugprnote2} We approximately divide the Hugoniot into three sections: structure at below $\sim$90 GPa is likely $\alpha$-B$_{12}$ or $\beta$, above $\sim$200 GPa is melt, and between 90--200 GPa is uncertain.}
\end{figure}

The Hugoniot curves thus obtained are shown with a
temperature-pressure $T-P$ diagram in Fig.~\ref{fig:hugTP}
and a pressure-density diagram in Fig.~\ref{fig:hugPr}.
Different single-phase Hugoniots are represented by different colored line-points. The temperature-pressure Hugoniots are compared with the equilibrium boron phase diagram based on~\cite{Solozhenko2013,Oganov2009}, 
and pressure-density Hugoniots are compared with shock compression experimental data from ~\cite{MarshLASL1980} and ~\cite{lepape2013}.

Our results for the liquid show that Hugoniot curves are insensitive to the simulation cell being used. This means once the sample is shocked to melt, the Hugoniot data does not exhibit any memory of the solid phase that existed right before the melt.  This effect is also evident in the warm dense matter Hugoniot measurements from Ref.~\cite{Zhang2018b}, where we find good 
agreement between the DFT-based Hugoniot and experiment.
In comparison, the Hugoniot data for solid phases could be characterized by discontinuities, as Fig.~\ref{fig:hugTP} shows.
If phase transitions occur in shock compressed boron at the same pressure-temperature conditions as those expected by the equilibrium phase diagram, the Hugoniot would have two major discontinuities, one at 15 GPa and the other at 80 GPa,  corresponding to the phase boundaries of $\alpha$-B$_{12}$$\leftrightarrow$$\gamma$-B$_{28}$ and $\gamma$-B$_{28}$$\leftrightarrow$$\alpha$-Ga, respectively. Following the DFT-based structural equilibrium picture, melting along the Hugoniot occurs for the $\alpha$-Ga phase at 2500--3500 K and 200 GPa, unless some other unknown structure stabilizes over $\alpha$-Ga at 100 GPa and 2000 K, which is not unlikely considering the fact that high temperature phases can be stabilized by high pressure in other elemental systems such as beryllium~\cite{LuPRL2017Be} and silicon~\cite{Paul2019}.

However, when comparing the pressure-density Hugoniots under the assumption that the phase changes follow the equilibrium phase diagram,
we find clear discrepancies between theory and experiment, as shown by the differences between the solid curves and black triangles in Fig.~\ref{fig:hugPr}. 
If a sample is shocked up to 80 GPa, the experimental data appear to follow the trend of our predicted Hugoniot for the $\alpha$-B$_{12}$ and $\beta$-B$_{106}$ phases, and clearly deviate from that of $\gamma$-B$_{28}$.
Above 80 GPa, all experimental Hugoniot data up to 112 GPa follow the trend of our predicted Hugoniot for $\alpha$-B$_{12}$ as well as that for $\gamma$-B$_{28}$ and probably also $\beta$-B$_{106}$, but do not show signatures of transition to the $\alpha$-Ga phase that would be associated with a down-jump in pressure and temperature and an up-jump in density.
These indicate that boron may remain in the same $\beta$ or $\alpha$-B$_{12}$ phase as its initial state when shock compressed to at least 80 GPa, instead of transforming into the $\gamma$-B$_{28}$ phase. 
Our simulations show temperature can significantly affect the stability of $\gamma$-B$_{28}$ and melting occurs upon very slight increase in temperature (Fig.~\ref{fig:eosgrid}).
Therefore, if the transformation to $\gamma$-B$_{28}$ happened at above 80 GPa, it would be associated with a jump in temperature and immediately followed by melting or transformation into some other solid structures.
In addition, our DFT-MD simulations show $\beta$-B$_{106}$ remains stable at $\sim$115 GPa and 600 K, and large atomistic displacement or structural instability occur when temperature exceeds 800--1000 K or pressure is higher than 130 GPa. We also find $\alpha$-B$_{12}$ remains stable at $\sim$133 GPa, and instability occurs at above 150 GPa, for 1000 K or lower temperatures. These set the upper bounds for $\beta$-B$_{106}$ or $\alpha$-B$_{12}$ samples to remain stable when boron is shocked to above 80 GPa.
With stronger shock to above 200 GPa and 3500 K, liquid boron is expected.
Temperatures along the shock Hugoniot are increasingly higher than cold compression along an ambient-temperature isotherm. The transformation kinetics is thus expected to be slower in room-temperature static compression experiments.

Our findings above on the phase transitions in shocked boron based on the EOS point of view are supported remarkably well by experiments.
As an example, in the diamond-anvil cell and x-ray diffraction experiments by Sanz {\it et al.}~\cite{SanzPRL2002amphous}, $\beta$-boron was the stable structure up to $\sim$100 GPa, at which point an amorphization structure was observed.
Similarly, explosive shock and diffraction experiments by Molodets and co-authors~\cite{Molodets2018,Molodets2017} did not detect the occurrence of $\gamma$-B$_{28}$ boron over the pressure range of 20--90 GPa. Instead, the $\beta$-B$_{106}$ phase was found to be stabilized until amorphization occured at 90--115 GPa.
In another experiment by Le Pape {\it et al.}~\cite{lepape2013}, the ionic structure factor $S(Q)$ at several scattering wavevectors ($Q$) was measured for boron laser-shocked to 100 GPa and 3.4 g/cm$^3$. By comparison with theoretical predictions of the solid and liquid Hugoniots, they attributed the results to a liquid state. However, there is no direct experimental evidence that allows one to distinguish these $S(Q)$ data from that of an amorphous solid, which is structurally similar to liquid boron~\cite{Krishnan1998prl}.

It is interesting to note that, the Hugoniot temperature of $\beta$-B$_{106}$ at 90 GPa is $\sim$600 K according to our calculations (Fig.~\ref{fig:hugTP}). 
The absence of $\gamma$-B$_{28}$ or $\alpha$-Ga phases in the shock experiments, together with findings in laser heated DAC experiments~\cite{Chuvashova2017} that it requires heating to $\sim$2000 K to make the phases out of $\beta$ boron, suggests an energy barrier of 0.05--0.17 eV between $\beta$ and $\gamma$/$\alpha$-Ga phases. The observed amorphization in megabar-pressure experiments~\cite{SanzPRL2002amphous,Molodets2017,Molodets2018} is likely a product of competition between the thermodynamic barrier that slows down the process of phase transformation and the decreased dynamic stability of $\beta$ boron at megabar pressures.

It is always the case that DFT-MD calculations are not perfect. For example, although the DFT-MD calculations accurately capture any anharmonic effect, treating atoms classically neglects the nuclear quantum effect. While this may be important in determining the absolute EOS accurately, the effect on the Hugoniot is likely smaller~\cite{Zhang2019BN} and could be largely canceled when comparing the differences between different phases. Therefore, we do not expect this effect to change our conclusions. However, considering the close relevance of nuclear quantum effects to structure searching and improving the equilibrium phase diagram, we plan to explore potential effects of nuclear quantum motion in a future publication.


\section{Conclusion}
In conclusion, by  extensive DFT-MD calculations of the EOS and Hugoniots of boron in various phases, we show that the phase transitions in boron under shock compression cannot be explained simply via comparison with the equilibrium crystaline phase diagram. The discrepancies between calculations and experimental data can be better reconciled by assuming over-pressurization of the $\alpha$-B$_{12}$/$\beta$ phase
prior to transition to the $\gamma$-B$_{28}$ phase,
and of the $\gamma$-B$_{28}$ phase prior to transition to $\alpha$-Ga or melting.
In addition, our computed melting temperatures are lower than expected and indicate the melting curve of boron has an anomaly and a negative Clapeyron slope at 100 GPa and 1500--3000 K.

Experiments at 100--200 GPa could test our predictions and are achievable using various DAC and shock experiments.
Dynamic DAC, laser shock compression with pre-compressed or pre-heated samples, decaying shock, and {\it in situ} diffraction could be very useful in clarifying the microscopic mechanism and property changes upon solid state phase transitions and melting of boron. 
Temporal velocity profiles in shock experiments with various setups will also provide important constraints  to quantify the kinetics of boron phase transitions by applying phase field models~\cite{Haxhimali2017}.

These results enable in-depth understanding of matter under shock compression, in particular the significance of compression-rate dependence of phase transitions and kinetic effects in experimental measurements.
Our theoretical findings, joint with recent experiments~\cite{McBride2019,Gorman2018,Coleman2019}, provide striking evidences showing that phase relations of materials under shock compression can deviate from those determined in equilibrium and raise questions about kinetics or non-equilibrium processes that materials may undergo during the timescale of the pressure loading.


\paragraph{Acknowledgments}  
This research was in part 
performed under the auspices of the U.S. Department of Energy by Lawrence Livermore National Laboratory under Contract No. DE-AC52-07NA27344.
Computational support was provided by LLNL high-performance
computing facility (Quartz). S.Z. was partially supported by the PLS-Postdoctoral Grant of LLNL. The authors acknowledge Dr. Alfredo Correa, Dr. Federica Coppari, and Prof. Eva Zurek for useful discussions.

This document was prepared as an account of work sponsored by an agency of the United States government. Neither the United States government nor any agency thereof, nor any of their employees,
makes any warranty, express or implied, or assumes any legal
liability or responsibility for the accuracy, completeness, or
usefulness of any information, apparatus, product, or process
disclosed, or represents that its use would not infringe privately owned rights. Reference herein to any specific commercial product, process, or service by trade name, trademark,
manufacturer, or otherwise does not necessarily constitute
or imply its endorsement, recommendation, or favoring by
the U.S. Government or any agency thereof. The views and
opinions of authors expressed herein do not necessarily state
or reflect those of the U.S. Government or any agency thereof, and shall not be used for advertising or product endorsement purposes.



\begin{thebibliography}{10}
\bibitem{initrhonote}
We choose $\beta$ boron with $\rho_0$=2.338~g/cm$^3$ here for the initial
  structure because it enables more reasonable comparison with most of the
  experimental data, which are from ~\cite{MarshLASL1980} using samples with
  the same initial density. The other two experimental data points
  from~\cite{lepape2013} have slightly higher initial density of
  $\sim$2.4~g/cm$^3$. We note that using $\alpha$-B$_{12}$ with
  $\rho_0$=2.46517 g/cm$^3$ as the initial structure, the Hugoniots are
  slightly different but our conclusions do not change.
\bibitem{hugprnote1}
We note that our predicted Hugoniot for the liquid are consistent with that by
  Plagemann {\it et al.} but different from those that reported in
  ~\cite{lepape2013}. However, complete details of the calculations and the
  experimental datapoint at 400 GPa are missing which makes further
  clarification of the underlying reasons impossible.
\end{thebibliography}

\end{document}